\documentstyle[aps,preprint]{revtex}
\tightenlines
\begin{document}
\draft
\title{THE NONLINEAR EFFECTS IN 2DEG  CONDUCTIVITY
INVESTIGATION BY AN ACOUSTIC METHOD.}

\author{ I.L.Drichko, A.M.Diakonov, V.D.Kagan, A.M.Kreshchuk,
T.A.Polyanskaya, I.G.Savel'ev, I.Yu.Smirnov, A.V.Suslov}

\address{A.F.Ioffe Physicotechnical Institute of RAS,
Polytechnicheskaya 26, 194021, St.Petersburg, Russia.}

\maketitle

\begin{abstract}

The parameters of two-dimensional electron
gas (2DEG) in a GaAs/AlGaAs heterostructure
were determined by an acoustical
(contactless) method in
the delocalized electrons region
($B\le$2.5T).

Nonlinear effects in Surface Acoustic Wave (SAW) absorption by 2DEG
are determined by the electron heating in the
electric field of SAW,
which may be described in terms of electron temperature $T_e$.
The energy relaxation time
$\tau_{\epsilon}$
is determined by
the scattering at piezoelectric potential of acoustic phonons
with strong screening.
At different SAW frequencies the heating depends
on the relationship between
$\omega\tau_{\epsilon}$ and 1 and is determined
either by the instantaneously changing wave field
($\omega\tau_{\epsilon}$$<1$),
or by the average wave power ($\omega\tau_{\epsilon}$$>1$).
\end{abstract}


Acoustical method employs a SAW
propagating on a surface of
piezoelectric substrate \cite{5,6}
(in our case $LiNbO_3$), which is accompanied by an electric field with
frequency $\omega=2\pi f$ equal to that of SAW,
penetrating in the 2DEG channel,
separated from the surface of a substrate by a vacuum gap $a$.
In experiments we measure the absorption of SAW by 2DEG.
Since the SAW absorption coefficient $\Gamma$
is determined by the conductivity
of 2DEG, quantizing of the electron spectrum in the magnetic field
should result in peculiarities
in absorption of SAW.

We studied the magnetic field $B$ dependence of
$\Gamma$
in GaAs/AlGaAs heterostructures
in $B$ up to 2.5T in $f$ range 30-210MHz
at $T$=1.4-4.2K in the linear regime
(input power $< 10^{-7}W$) and at $T$=1.5K for
different SAW intensities.
From previous studies \cite{7} on the same samples we obtain
the carrier density  $n = 6.7\cdot 10^{11} cm^{-2}$ and
the mobility $\mu= 1.28\cdot 10^{5} cm^2/Vs$ at $T$=4.2 K (DC-data).

Curve $1$ on figure 1 illustrates
the $\Gamma$ dependences on the magnetic
field $B$ at $T=4.2K$.
The maxima of $\Gamma$, as and minima of resistivity in a case of
direct current measurements,
are equidistant in inverse magnetic field.
That allows one to find the carrier density, by the standard method,
which yields $n=7\cdot 10^{11} cm^{-2}$.
From $\Gamma(B)$, $\sigma_{xx}(B)$ was determined
using the formula for $\Gamma(\sigma_{xx})$ of \cite{Kag}, and
using a value of $a$ obtained in  a way presented
in \cite{7,8}.

It was shown in \cite{7} that at $B<2.5T$,
$\sigma_{XX}^{AC}$ derived from the direct current measurements
and $\sigma_{XX}^{DC}$ derived from SAW absorption are equal,
and we beleive electrons to be delocalized there.
In this case one can determine another parameters of
the 2DEG in the heterostructure.

Curve $2$ of figure 1 presents the dependence $\sigma_{xx}$ on $B$
derived from dependence of $\Gamma(B)$.

In accordance with
the Ando theory \cite{Ando}
$\sigma_{xx}= \sigma_{xx}^*+\sigma_{xx}^{osc}$,
where $\sigma_{xx}^*=\sigma_0/(1+\omega_c^2\tau_p^2)$ is the
classical Drude conductivity,
$\tau_p$ is the transport lifetime,
$\omega_c=eB/m^*c$ is the cyclotron frequency,
$m^*$ is the effective mass, and $\sigma_0$ is the zero-field conductivity.
The oscillatory term is \cite{Ando,Coleridge}:
\begin{eqnarray}
\sigma_{xx}^{osc}\propto \sigma_{xx}^* D(X_T)exp(-\pi/\omega_c\tau_q)
cos(2 \pi E_F/\hbar\omega_c-\pi)
=\Delta\sigma_{xx}cos(2 \pi E_F/\hbar\omega_c-\pi)/2,\label{eq:sigmaosc}
\end{eqnarray}
where $D(X_T)=X_T/sinh(X_T)$, $X_T=2\pi^2T/\hbar\omega_c$,
is determined
by the temperature broadening of the Fermi level,
$E_F$ is the Fermi energy,
$\tau_q$ is the quantum lifetime,
which is a measure of the collisonal broadening of
the Landau levels at the value of $A=\hbar/2\tau_q$,
and $\Delta\sigma_{xx}$ is the oscillation amplitude.

From the slope of $\sigma_{xx}^*(1/B^2)$ one obtaines
$\mu_0$ at $B=0$.
The mobility $\mu_0$ appeared to be equal to
$(1.1\pm0.1) 10^5 cm^2/V\cdot s$
(15 percent difference from the Hall mobility)
and does not depend on temperature.

From the analysis of the maximal amplitude $\Delta\sigma_{xx}$ value (fig.1)
one can get the classical-to-quantum scattering time ratio
$\tau_p/\tau_q$, which is equal to $5.5\pm0.5$
and corresponds to a case of dominance of
screened-Coulomb scattering of electrons
from ionized charge centers \cite{harrang}.
The Dingle temperature is
$T^*=\hbar/2\pi\tau_q=1.5K$.
In accordance with \cite{Sarma} using the known ratio $\tau_p/\tau_q$
and the carrier sheet density one can estimate the spacer thickness, which
takes a value of $\sim30\AA$.

Let's consider the nonlinear effects at relatively low $B$.
The dependences  of $\Gamma$ on
$T$ and SAW intensity
were obtained from  the experimental curves of the type
presented in Fig.1.
Fig.2a illustrates the temperature dependence
of $\Delta\Gamma=\Gamma_{MAX}-\Gamma_{MIN}$, where $\Gamma_{MAX}$ and
$\Gamma_{MIN}$ are the values of an adjacent minimum and maximum of $\Gamma$,
measured
in a linear regime at a frequency of 150 MHz.
Fig.2b presents the  $\Delta\Gamma$ dependence on the output power
$P$ of the RF source of SAW
at 150 MHz at $T$=1.5 K.
It is seen, that
 $\Delta\Gamma$ decreases with the $P$ and $T$ increase.

It's quite natural to consider the nonlinearity mechanism, which
was investigated earlier on these structures using
direct current
 \cite{9}, where
it has been shown that the dependence of the resistivity on the current
density can be explained by the 2DEG heating.

For a description of the electron gas heating using temperature
$T_e$, which differs from the lattice temperature
$T_0$, the condition $\tau_p << \tau_{ee} << \tau_{\epsilon}$
is to be met
($\tau_{ee}$, $\tau_{\epsilon}$ are
the electron-electron collision time
and energy relaxaton time respectively). $\tau_p$
can be derived from the value of $\mu_0$  and is equal to
 $\tau_p =5\cdot 10^{-12}s$.
In \cite{Non97} we obtained $\tau_{ee}=5\cdot 10^{-10}s$,
 $\tau_{\epsilon} \simeq 3.3\cdot 10^{-9}s$.

By analogy with \cite{9} one can determine $T_e$-
the temperature of 2DEG- comparing the dependence of
 $\Delta\Gamma$   on $P$ with its temperature dependence .

In order to find the energy losses $Q$,
we have made the following calculations:
the electric field $E$, corresponding to a SAW propagating
in a piezoelectric, and penetrating in a 2D system, is:
\begin{eqnarray}
\  |E|^2=K^2\frac{32\pi }{\nu}(\epsilon_1+\epsilon_0)b
\frac{kexp(-2ka)}
{1+[(4\pi \sigma_{xx}/\epsilon_s v)c(k)]^2 }W,\label{eq:2}
\end{eqnarray}
where
$K^2$ is the electromechanical coupling coefficient, $v$, $k$ are
the SAW velocity and the wave vector;
$\epsilon_0$ ,
$\epsilon_1$ , $\epsilon_s$ are the dielectric constants of
vacuum, the piezoelectric and the sample respectively,
$W$ is the
input SAW power in the sample
per sound track width.
The functions $b$ and $c$ are complex functions of
$\epsilon_0$, $\epsilon_1$, $\epsilon_s$,
 $k$, $a$.
The energy losses per one electron
 $Q=e\mu E^2$. Multiplying
the both parts of Eq.(2) by $\sigma_{xx}$, one obtains
$Q=4W\Gamma/n$.

We have invesigated the dependencies $Q=f(T_e^3-T_0^3)$, which in accordance
with the results of \cite{9} obtained for a DC, should correspond to
the energy relaxation of electrons on
piezoelectric potential of acoustic phonons
($PA$-scattering) in case of weak screening \cite{10}.
However, we find that our
experimental curves can be fitted  better by
$Q=A_5(T_e^5-T_0^5)$ which corresponds to
the case of $PA$-scattering with strong screening
\cite{10}.
This is also in better accordance with
our experimental situation, because the strong screening condition is
fulfilled for the studied sample.
Figure 3 shows the experimental points and theoretical dependencies
$Q(T_e^5-T_0^5)$
for $f$=30 and 150 MHz with $A_5=3\pm0.5eV/sK^5$
and $4.1\pm0.6eV/sK^5$, respectively.

Besides, it should be noticed that $A_5$ takes different values
for different frequencies.
This difference is
associated probably due to the fact that
in the case of 150MHz
$\omega\tau_{\epsilon}>1$ the heating
is determined by average wave power, whereas at $f$=30MHz and
$\omega\tau_{\epsilon}<1$
the 2DEG heating is
determined by the instantaneously changing wave field.
That can lead to the different heating degree
for the same energy losses.

The work was supported by the
RFFI N95-02-04066-a and MINNAUKI N97-1043 grants.

\begin{figure}
\caption{The experimental dependences of
the absorption coefficient $\Gamma$ ($1$)
and of the 2DEG conductivity $\sigma_{xx}$($2$)
on magnetic field at $T=4.2K$ at frequency 30MHz.}
\end{figure}
\begin{figure}
\caption{The $\Delta\Gamma$ dependence on temperature $(a)$ and on
power $(b)$ at 150MHz for different magnetic fields $B$:
1-1.75T, 2-1.5T, 3-1.41T.}
\end{figure}
\begin{figure}
\caption{The dependence of the energy losses $Q$ on the temperature
$T_e$ for two frequencies (1-150MHz, 2-30MHz).}
\end{figure}

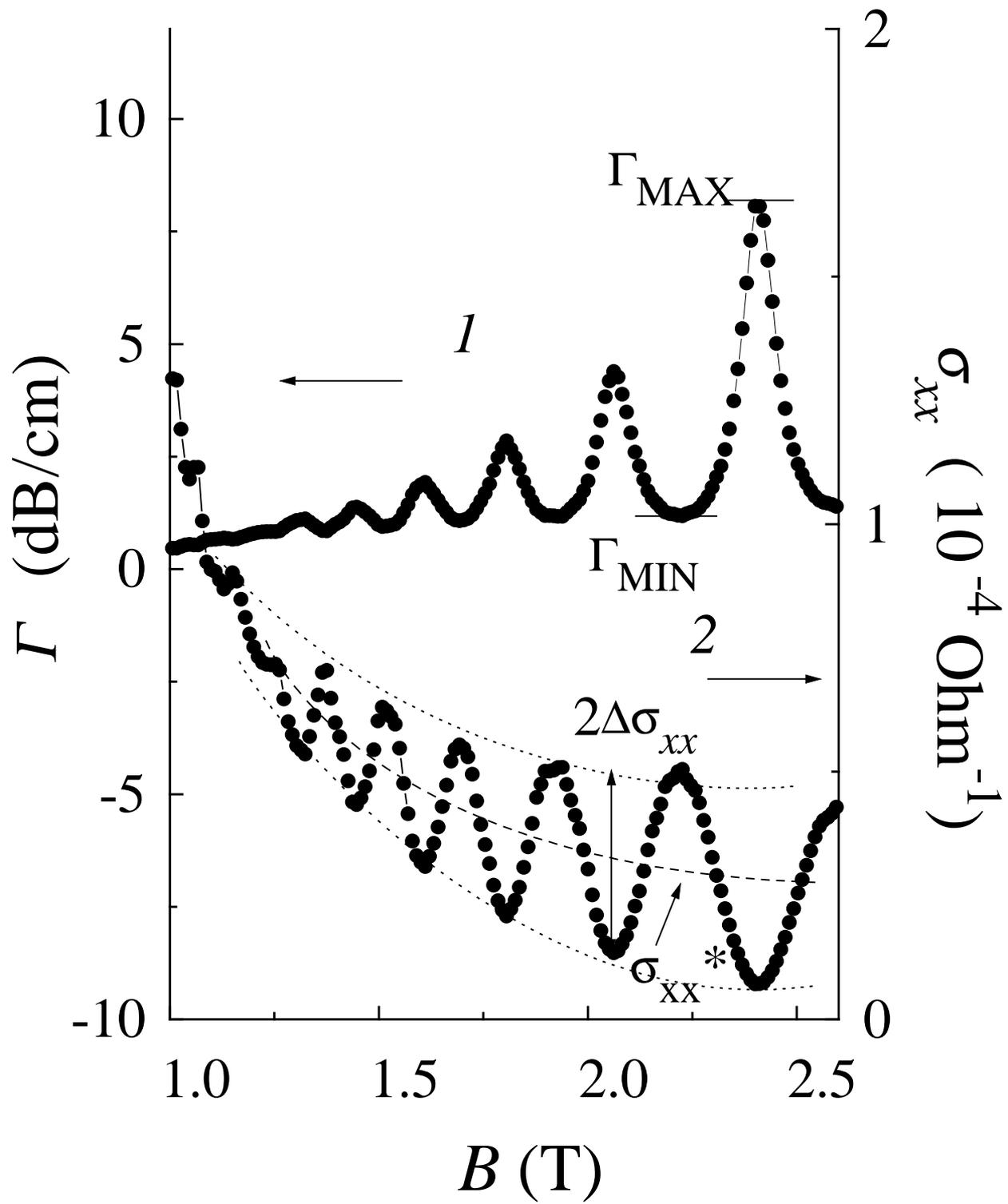
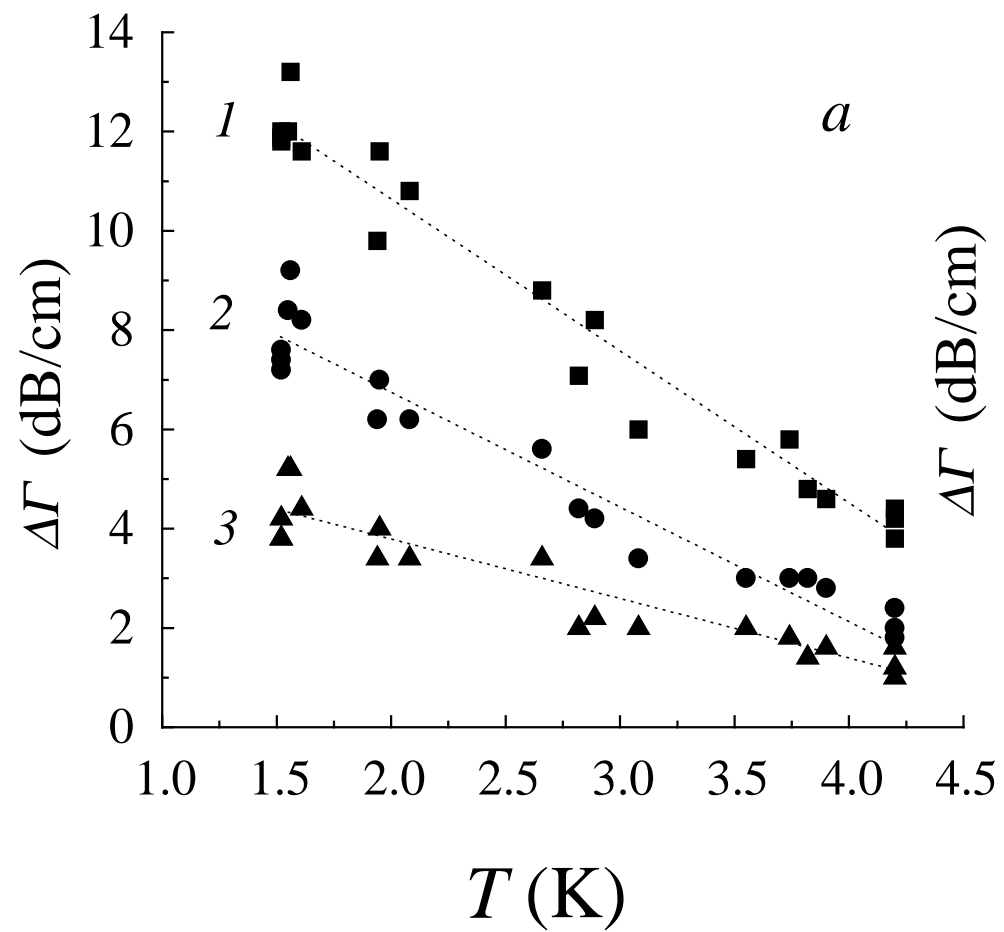
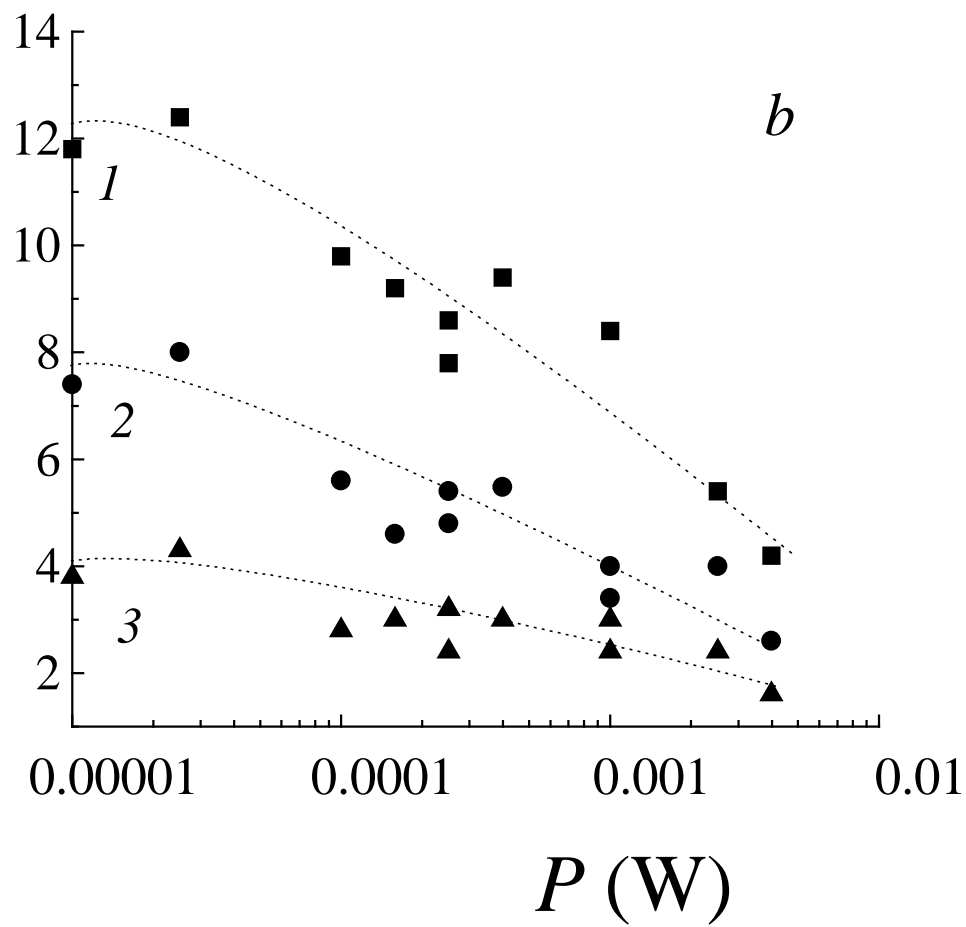
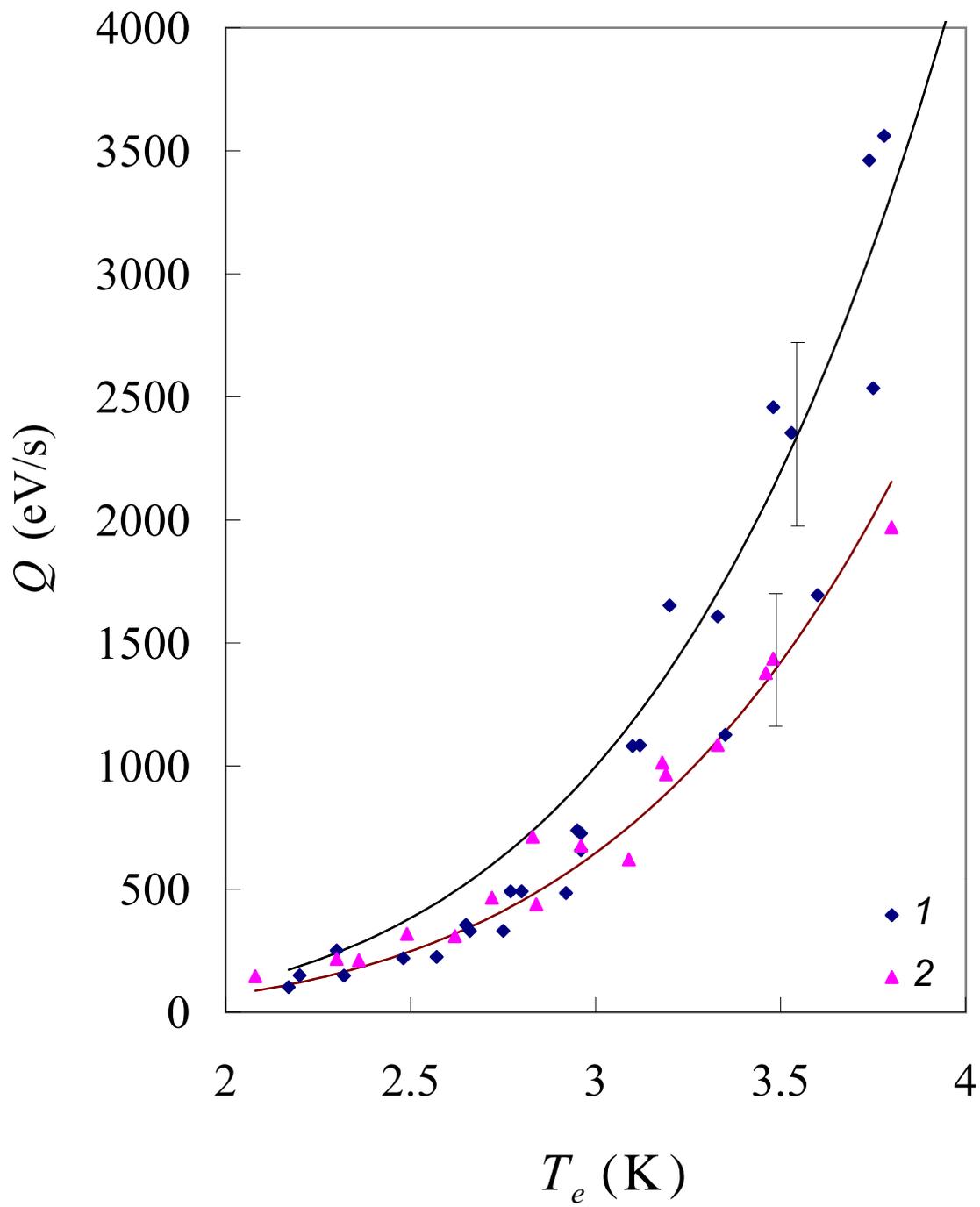
\end{document}